# WST – Widefield Spectroscopic Telescope: addressing the instrumentation challenges of a new 12m class telescope dedicated to widefield Multi-object and Integral Field Spectroscopy

David Lee[a], Joel D. R. Vernet[b], Roland Bacon[c], Alexandre Jeanneau[c], Ernesto Oliva[d], Anna Brucalassi[d], Andrea Tozzi[d], José A. Araiza-Durán[d], Andrea Bianco[e], Jan Kragt[f], Ramon Navarro[f], Bianca Garilli[g], Kjetil Dohlen[h], Jean-Paul Kneib[i], Ricardo Araujo[i], Maxime Rombach[i], Eloy Hernandez[j], Roelof S. de Jong[j], Andreas Kelz[j], Stephen Watson[a], Tom Louth[a], Ian Bryson[a], Elizabeth George[b], Norbert Hubin[b], Julia Bryant[k], Jon Lawrence[l]
.
[a] UK Astronomy Technology Centre (United Kingdom), [b].European Southern Observatory (Germany), [c].Observatoire de Lyon, [d] INAF - Osservatorio Astrofisico di Arcetri (Italy), [e] INAF - Osservatorio Astronomico di Brera (Italy), [f] ASTRON (Netherlands), [g] INAF - Istituto di Astrofisica Spaziale e Fisica cosmica Milano (Italy), [h] Laboratoire. d'Astrophysique de Marseille (France); [i] Ecole Polytechnique Fédérale de Lausanne (Switzerland), [j] Leibniz-Institut für Astrophysik Potsdam (Germany), [k] Sydney Institute for Astronomy, The University of Sydney (Australia), [l] Macquarie University. (Australia).

**ABSTRACT**

WST – Widefield Spectroscopic Telescope: We summarise the design challenges of instrumentation for a proposed 12m class Telescope that aims to provide a large (>2.5 square degree) field of view and enable simultaneous Multi-object (> 20,000 objects) and Integral Field spectroscopy (inner 3x3 arcminutes field of view), initially at visible wavelengths. For the MOS mode, instrumentation includes the fiber positioning units, fiber runs and the high (R~40,000) and low (R~3,000 - 4,000) resolution spectrographs. For the MUSE like Integral Field Spectrograph, this includes the relay from the Telescope Focal Plane, the multi-stage splitting and slicing and almost 150 identical spectrographs. We highlight the challenge of mass production at a credible cost and the issues of maintenance and sustainable operation.

## 1. INTRODUCTION

The Widefield Spectroscopic Telescope (WST) is an ambitious and unique facility that will allow breakthrough science in many areas of modern astrophysics. The motivation and science drivers for WST are described elsewhere in the proceedings of this conference [1]. WST will be a 12 m class wide-field spectroscopic survey telescope with simultaneous operation of both Integral Field Spectroscopy (IFS) and Multiple-Object Spectroscopy (MOS) observing modes. In addition, the MOS capability will include both low- and high-resolution modes. The top-level requirements for WST's telescope and instrumentation are listed in Table 1. The initial set of instruments are specified to operate over the wavelength range 370 nm to 970 nm, but the telescope will be designed to operate over a broader wavelength range of 350 nm to 1600 nm. This will enable a future upgrade path to include near-infrared instruments.

A computer aided design model of the WST building, telescope, and instrumentation is shown in Figure 1. The telescope is a Cassegrain design, with a three-element field corrector providing a wide field of view of 2 degrees in diameter. Full details of the telescope optical design, including the atmospheric dispersion compensation, can be found in [2]. The ~F/3.3 MOS focal surface is located approximately 1.6 m behind the last element in the field corrector and is approximately 1.4 m in diameter. The MOS focal plate will be equipped with approximately 20,000 to 22,000 robotic fiber positioners. The low- and high- resolution MOS spectrographs will be located on the azimuth floor, whilst the IFS is in the pier, below the telescope.

Table 1. WST top-level requirements.

| | |
|---|---|
| Telescope Aperture | 12 m, seeing limited |
| Telescope FoV | 3.1 deg$^2$ |
| Tel. Spec Range | 350 -1600 nm |
| MOS LR Multiplex | 20,000 |
| MOS LR Resolution | 3,000-4,000 |
| MOS LR Spec Range | 370-970 nm (simultaneous) |
| MOS HR Multiplex | 2,000 |
| MOS HR Resolution | 40,000 |
| MOS HR Spec Range | 370-970 nm (3-4 regions) |
| IFS FoV | 3x3 arcmin$^2$ |
| IFS Resolution | 3,500 |
| IFS Spec Range | 370-970 nm (simultaneous) |
| IFS Patrol Field | 13 arcmin (diameter) |
| MOS & IFS simultaneous operation | |
| ToO implemented at telescope and fibre level | |

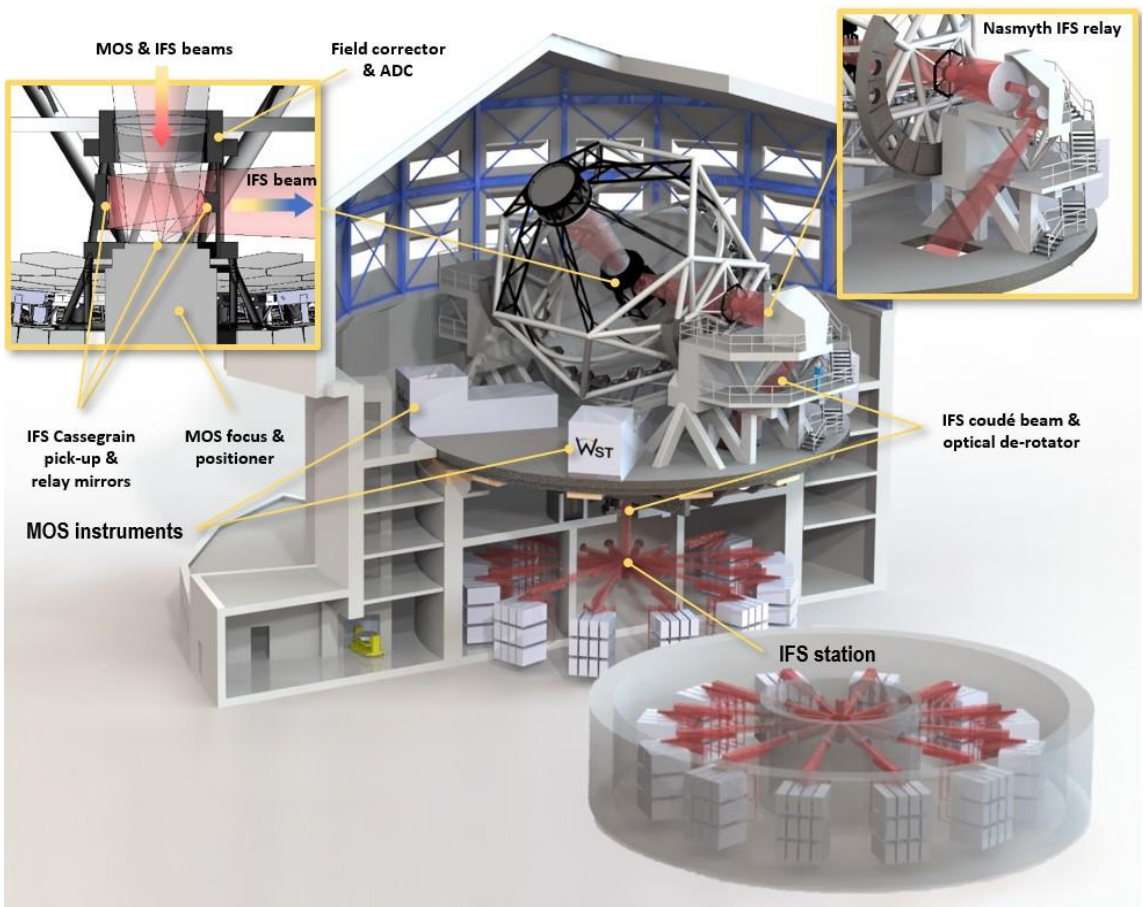

Figure 1. WST overall layout.

The field of view for the IFS is extracted from the center of the MOS focal surface. A fold mirror, shown in the top-left inset in Figure 1, directs light to an optical relay, located between the field corrector and the MOS focal surface, which sends the IFS beam horizontally towards the Nasmyth platform. A further set of mirrors mounted on the Nasmyth platform form an intermediate image 13 arcminutes in diameter, of which a 3 arcminute by 3 arcminute sub-field is selected for further propagation to the IFS Coudé focal plane. A further mirror system forms the Coudé relay, which reimages the Nasmyth focus to the Coudé focus, and includes a field derotation system. The final Coudé image, formed underneath the telescope in the area labelled "IFS station", is F/35.

WST will be equipped with a large flat-field screen mounted inside the dome. This will be illuminated with various calibration light sources to provide accurate and efficient calibration of the MOS and IFS instruments.

Below the telescope azimuth floor is a services level that contains electronics cabinets, cables, cooling pipes, etc. This level also accommodates some of the Coudé focus relay mirrors which must co-rotate with the telescope. The IFS field derotation system is housed within the central pier of the telescope.

The small yellow object in the lower left of the WST building, Figure 1, indicates a handling trolley for the movement of the low-resolution MOS spectrograph blocks. Similar trolleys will be used in other areas, for example for movement of telescope primary mirror segments for re-coating. The layout of the building needs to be optimized for handling and maintenance given the size and complexity of the instrumentation suite. Primary mirror segments will be removed and replaced by a deployable crane.

## 2. INTEGRAL FIELD SPECTROGRAPH

The WST integral field spectrograph will be in the stationery Coudé room at the base of the telescope structure, as shown in Figure 1. On reaching the IFS the focal plane is split into 12 beams using a first stage field splitter mirror array, the conceptual design of which is shown in Figure 2 (left). The twelve second-stage beams can be seen in Figure 1, lower-right, in the central area of IFS station's circular room. Each of these beams illuminates a second stage field splitter, shown in Figure 2 (right) which produces twelve further beams, as shown in Figure 3, that illuminate a block of twelve integral field spectrographs. Twelve blocks of twelve IFS spectrographs are arranged symmetrically within the IFS station. The overall layout of the IFS is based on the MUSE instrument in operation on the European Southern Observatory's Very Large Telescope [3]. The electronics and cooling systems for 144 integral field spectrographs will be extremely complex and require significant volume. Technology development studies are planned to investigate options to miniaturize and simplify these areas, such as the use of external versus integrated detector controllers.

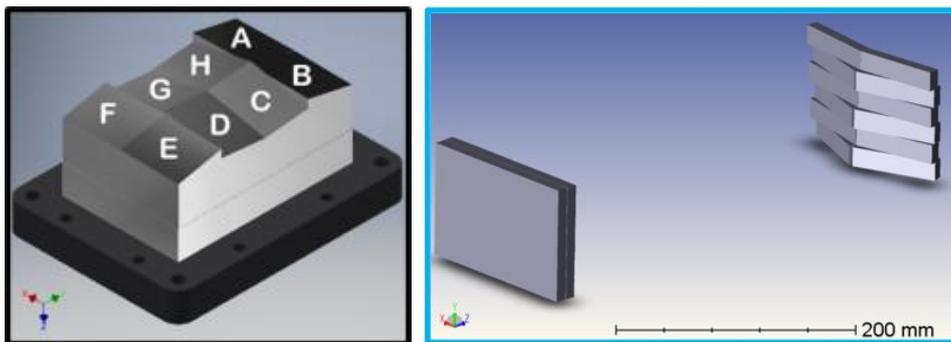

Figure 2. Conceptual design of the first stage field splitter showing eight mirrors (left) and second stage field splitter with twelve lenses and matching mirrors (right).

Figure 4 shows a preliminary optical design of an integral field spectrograph with two wavelength channels. The wavelength split is achieved with a dichroic beam-splitter filter. The optical design is all refractive with a three-element collimator lens, a volume phase holographic grating disperser, and a four-element camera lens. The optical design assumes the use of a curved detector, which helps to simplify the design, hence less optical surfaces are needed to achieve the image quality requirements. The design of the spectrograph is based on that previously described in [4].

A significant challenge for the design of the IFS will be to enable mass manufacturing of components to minimize cost.

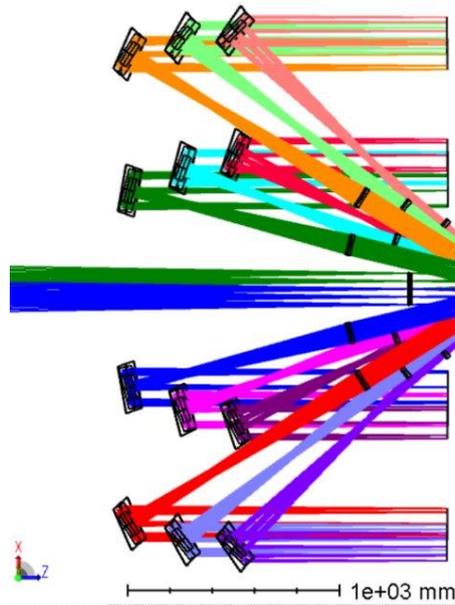

Figure 3. Optical design concept for the second stage splitter and relay system to the bank of twelve spectrographs. Light entering center left first passes through the field splitter lens array and then reflects from the second stage field splitter. The beam then passes through a second lens array before being reflected towards the spectrograph.

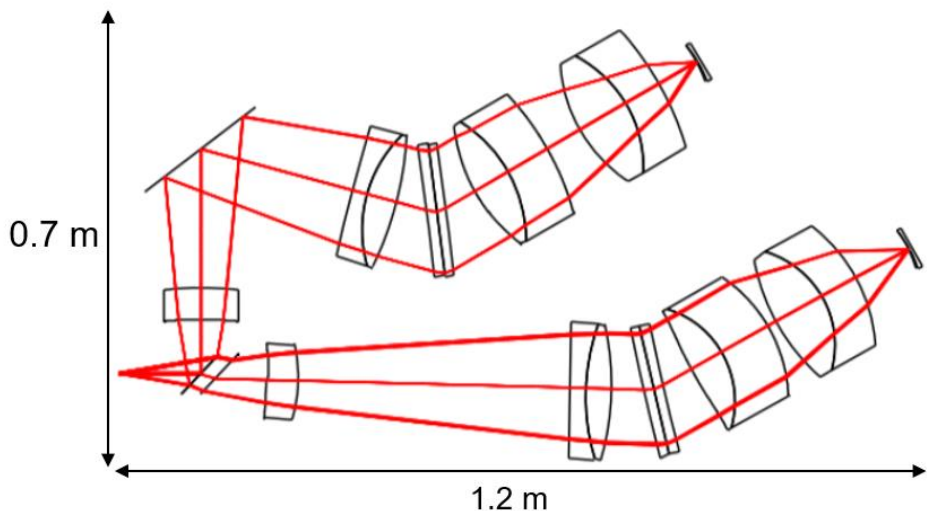

Figure 4. Optical ray-trace diagram of a two-wavelength-channel integral field spectrograph.

## 3. FIBER POSITIONER

The design of the fiber positioning system for WST is being developed by three collaborating organizations: Ecole Polytechnique Federale de Lausanne (EPFL), Leibniz Institute for Astrophysics Potsdam (AIP), and United Kingdom Astronomy Technology Centre (UKATC). Technology development studies are ongoing looking at three types of grid-based fiber positioning technology: Phi-Theta actuators, with two rotating axes, Phi-R actuators, with one rotating and one linear axis, and a new solution based on three actuators. The use of Phi-Theta fiber positioners is well established and already in use in instruments such as DESI [5] and PFS [6]. A range of Phi-Theta actuators has been developed by EPFL and are shown in Figure 5. Test results for these fiber positioners are reported in [7].

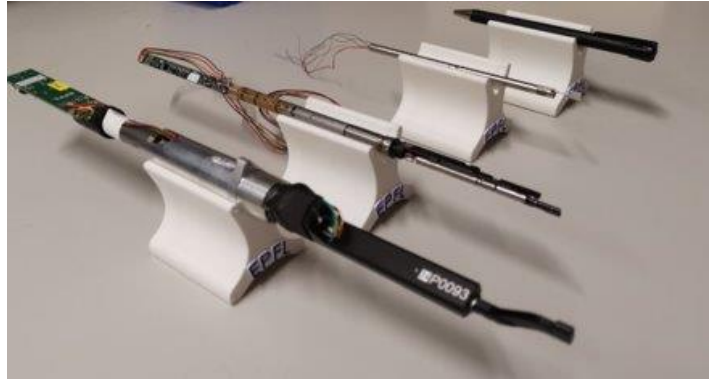

Figure 5. Examples of fiber positioners developed at EPFL. The largest has 22.4 mm pitch and the smallest has 5.5 mm pitch.

Phi-R fiber positioners are under development and are described in [8] and [9]. The new fiber positioning concept being developed by AIP [10] is based around a tripod actuator mechanism and potentially provides large patrol field (>30 mm), whilst having small pitch (< 7 mm).

A preliminary concept for layout of the MOS focal plate is shown in Figure 6. In this example groups of 75 fiber positioners are assembled into modular triangular holders, or rafts, with 324 rafts being assembled to fill the field of view of WST. This concept is based on one already developed for MegaMapper [11] with designs for WST reported in [12].

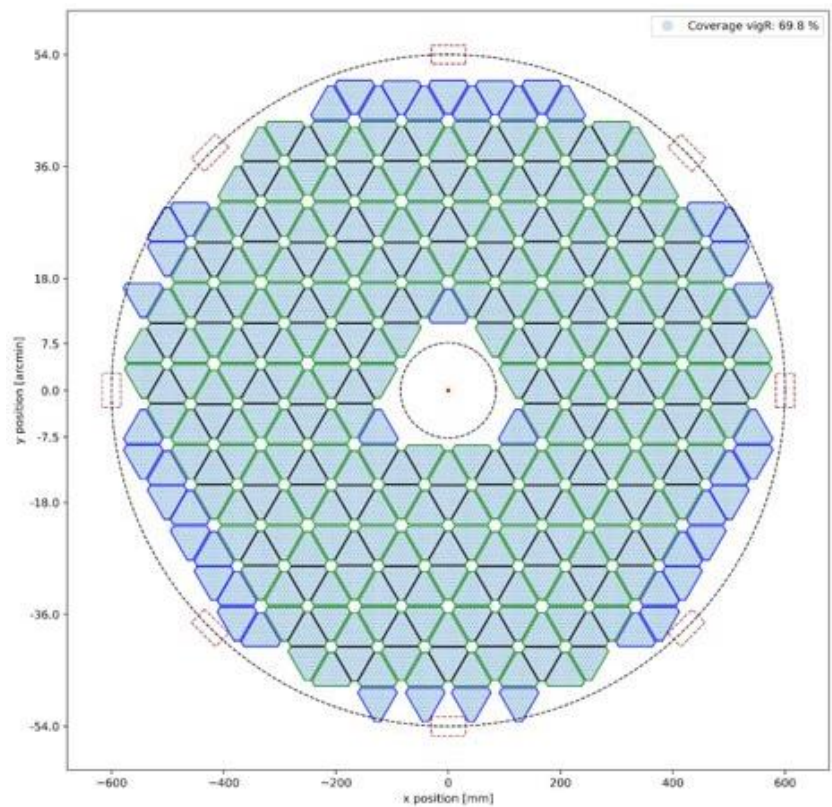

Figure 6. Example focal plane layout for an outside diameter of 1.2 m and a central obscuration of 0.17 m. Each triangle represents a module containing 75 fiber positioning robots. There are 324 modules giving a total of 24,300 fiber positioners.

# 4. MULTIPLE-OBJECT SPECTROGRAPH

WST will have two sets of multiple-object spectrographs to cover low- and high-resolution observations. Several design options for the low-resolution MOS have been investigated based on existing designs such as 4MOST [13], MOONS [14], and the IFS design shown in Figure 4.

An example preliminary design concept for the low-resolution MOS spectrograph, based on the IFS design, is shown in Figure 7. The optical configuration consists of a curved slit, three element refractive collimator, transmissive disperser, four element refractive camera, and a curved detector. The design shown has two wavelength channels, indicated by red and blue rays, but a third wavelength channel can be added by using an additional dichroic beam-splitter. The ray path for the third green channel is indicated by the green arrow. Note the addition of the third green channel would increase the space envelope of the spectrograph.

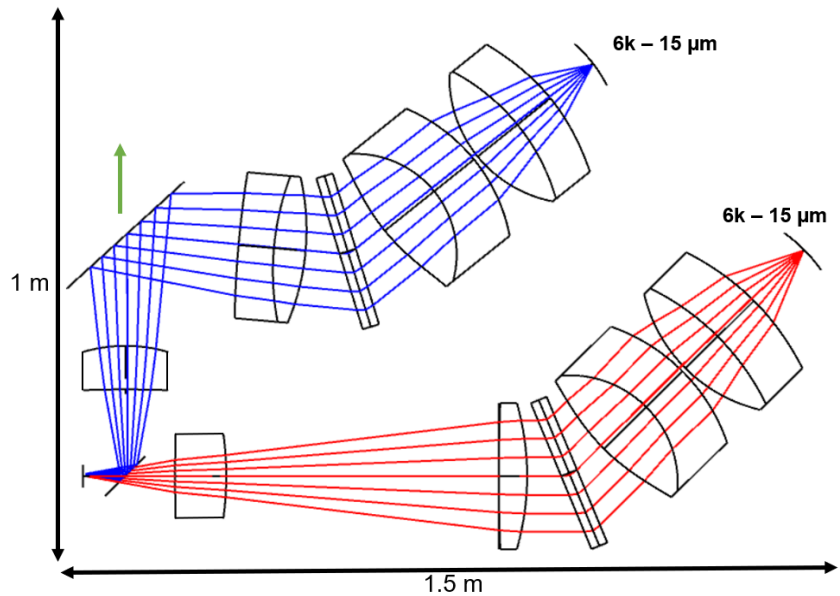

Figure 7. Optical ray-trace diagram of a two or three-channel multiple-object spectrograph

A preliminary design concept for the high-resolution MOS spectrograph is shown in Figure 8. The optical configuration consists of a curved slit, collimator mirror with refractive field corrector, transmissive disperser, catadioptric F/1.1 camera, and a detector mosaic with a two by two array of 9k × 9k pixel detectors. The camera design is a scaled-up version of the successful MOONS camera [14]. There are four separate spectrographs to cover four wavelength channels. Wavelength splitting is performed in the fiber link. To keep the fiber length to a minimum the high-resolution MOS spectrograph will be located next to the telescope on the azimuth floor, as seen in Figure 1.

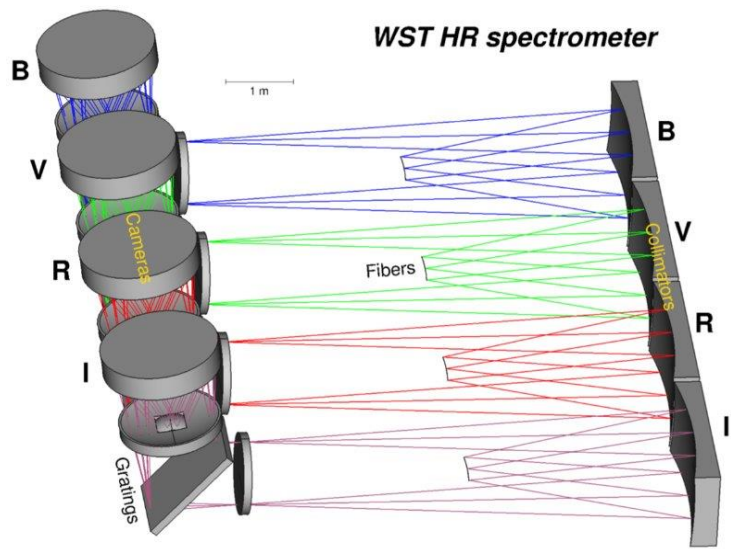

Figure 8. Design concept for the WST high-resolution spectrograph.

## 5. TECHNOLOGY DEVELOPMENT

Over the timescale for the development of WST there may be significant changes in the available technology needed for the construction of the instrumentation. Three key technology development areas have been identified and these will be investigated for WST:

1. Detector technology, CCD versus CMOS, and curved detectors.
2. Disperser technology; availability of larger gratings.
3. Fiber connectors.

Each of the three types of spectrographs require different format detectors: 4k × 4k, 15 µm pixels for the IFS, 6k × 6k, 15 µm pixels for the low-resolution MOS, and 9k × 9k, 10 µm pixels for the high-resolution MOS. In addition, the use of curved detectors for WST is expected to reduce the number of optical elements in the spectrograph designs. The top-level requirements for the detectors are therefore as follows: radius of curvature less than 250 mm, pixel size 10 – 15 µm, and linear size (along edge) 60 – 90 mm. As large numbers of detectors are needed for WST a custom mass production run becomes a possibility. The use of CMOS sensors provides advantages such as lower read noise and cost. The development of curved CMOS sensors, as shown in Figure 9, is already being investigated for other projects [15] and will be further investigated in future by the WST consortium.

The large number of detectors needed for the IFS and MOS will each require an enclosure (vacuum vessel) and cooling system. As well as the enormous complexity of this cooling system it is also likely to be the most significant user of energy by WST during operation. The WST consortium is investigating energy efficient cooling systems as part of the sustainability plan.

The low-resolution MOS and IFS will likely be designed to use high efficiency volume phase holographic gratings. The high-resolution spectrograph baseline solution is to use etched lithographic gratings. A new facility to produce volume phase holographic gratings is being prepared by INAF and is expected to be able to produce gratings up to 450 mm in diameter [16].

It is clear the joining and routing of ~22,000 optical fibers for WST will be another significant challenge. High efficiency fiber connectors will be needed that potentially make hundreds of fiber-to-fiber connections in a single device. These types of devices are the subject of on-going research activity within the WST consortium partners.

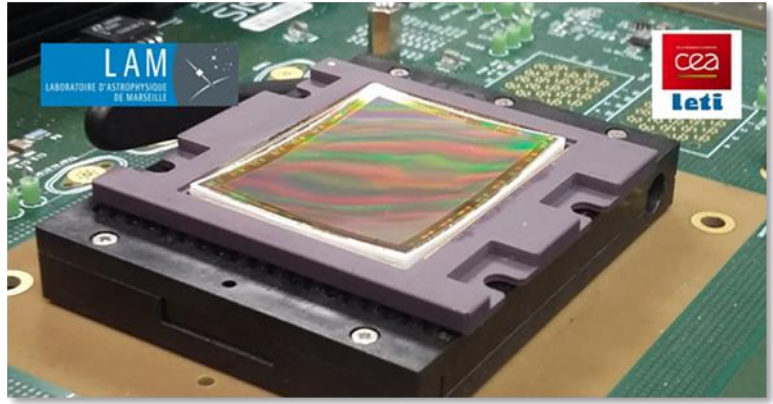

Figure 9. Picture of a curved detector being developed by CNRS-LAM and CEA-LETI.

## 6. SUMMARY & CONCLUSIONS

This paper describes the Wide-Field Spectroscopic Telescope and the preliminary design concept for the instrumentation. It has been shown that it is possible to design and manufacture the WST instrumentation using a combination of existing technology and new technology already under development.

As of 2024 more information can be found at the WST website: https://www.wstelescope.com/

## ACKNOWLEDGEMENTS

R. Bacon, P. Dierickx thanks CNRS INSU CSAA and ANR-AA-MRSE-2023 for their support. E. Oliva, A. Brucalassi, A. Tozzi, J. A. Araiza-Durán, A. Bianco and B. Garilli acknowledge support from INAF - Astrofisica Fondamentale 2023 Large Grant "The WST".